\documentclass[conference]{IEEEtran}

\usepackage[left=0.625in, right=0.625in, bottom=1in, top=0.7in]{geometry}

\usepackage{placeins}

\usepackage[nopostdot,nogroupskip,nonumberlist]{glossaries}

\newacronym{RSSI}{RSSI}{Received Signal Strength Indicator}

\newacronym{WEP}{WEP}{Wired Equivalent Privacy }
\newacronym{WPA}{WPA}{Wireless Protected Access}
\newacronym{PMK}{PMK}{Pairwise Master Key}
\newacronym{PTK}{PTK}{Pairwise Transition Keys}
\newacronym{GTK}{GTK}{Group Temporal Key}
\newacronym{EAP}{EAP}{Extensible Authentication Protocol}
\newacronym{RRM}{RRM}{Radio Resource Management}
\newacronym{BSSTM}{BSSTM}{BSS Transition Management}
\newacronym{BSS}{BSS}{Basic Service Set}
\newacronym{BSSID}{BSSID}{BSS Identifier}
\newacronym{CSA}{CSA}{Channel Switching Announcement}
\newacronym{EDCF}{EDCF}{Enhanced Distributed Coordination Function}
\newacronym{AIFS}{AIFS}{Arbitration Inter-Frame Space}

\newacronym{PSM}{PSM}{Power Save Mode}
\newacronym{APSD}{APSD}{Adaptive Power Save Delivery}

\newacronym{DBS}{DBS}{Disassociation-Based Steering}

\newacronym{SDN}{SDN}{Software Defined Networking}

\newacronym{VNF}{VNF}{Virtual Network Function}
\newacronym{NNF}{NNF}{Native Network Function}

\newacronym{PCP}{PCP}{Priority Code Point}

\newacronym{NAT}{NAT}{Network Address Translation}

\newacronym{NUMA}{NUMA}{Non-Uniform Memory Access}

\newacronym{CPE}{CPE}{Customer Premise Equipment}

\newacronym{DSCP}{DSCP}{Differentiated Services Code Point}

\newacronym{vSG}{vSG}{Virtual Subscriber Gateway}

\newacronym{RGW}{RGW}{Residential Gateway}
\newacronym{vRGW}{vRGW}{Virtual RGW}

\newacronym{VF}{VF}{Virtual Function}
\newacronym{PF}{PF}{Physical Function}

\newacronym{ARP}{ARP}{Address Resolution Protocol}

\newacronym{DTIM}{DTIM}{Delivery Traffic Indication Message}
\newacronym{AP}{AP}{Access Point}
\newacronym{TWT}{TWT}{Target Wake-up Time}
\newacronym{OPS}{OPS}{Opportunistic Power Save}
\newacronym{PCF}{PCF}{Point Coordination Function}
\newacronym{PIFS}{PIFS}{PCP Inter Frame Space}
\newacronym{DIFS}{DIFS}{Distributed Inter Frame Space}
\newacronym{BSR}{BSR}{Buffer Status Report}
\newacronym{OCW}{OCW}{OFDM Contention Window}

\newacronym{QTP}{QTP}{Quite Time Period}
\newacronym{AU}{AU}{Airtime Utilization}

\newacronym{LVAP}{LVAP}{Light Virtual AP}
\newacronym{EDCA}{EDCA}{Enhanced Distributed Channel Access}

\newacronym{HCCA}{HCCA}{Hybrid Controlled Channel Access}

\newacronym{DC}{DC}{Datacenter}
\newacronym{CO}{CO}{Central Office}

\newacronym{CORD}{CORD}{Central Office Re-architected as a Datacenter}
\newacronym{MEC}{MEC}{Mobile Edge Computing}
\newacronym{ASP}{ASP}{Application Service Provider}
\newacronym{ES}{ES}{Edge Service}
\newacronym{NSP}{NSP}{Network Service Provider}

\newacronym{NV}{NV}{Network Virtualization}
\newacronym{NFV}{NFV}{Network Function Virtualization}
\newacronym{NFVI}{NFV}{NFV Infrastructure}
\newacronym{MANO}{MANO}{Management and Orchestration}

\newacronym{vNIC}{vNIC}{virtual NIC}
\newacronym{pNIC}{pNIC}{physical NIC}

\newacronym{TCAM}{TCAM}{Tenary Content Addressable Memory}
\newacronym{RSS}{RSS}{Receive Side Scaling}

\newacronym{DPDK}{DPDK}{the Data Plane Development Kit}
\newacronym{UIO}{UIO}{User-space I/O}

\newacronym{EMC}{EMC}{Exact Match Cache}

\newacronym{SR-IOV}{SR-IOV}{Single Root-Input/Output Scaling}
\newacronym{PMD}{PMD}{Poll Mode Driver}

\newacronym{TSS}{TSS}{Tuple Space Search}
\newacronym{dpcls}{dpcls}{data path classifier}

\newacronym{AFDX}{AFDX}{avionics full-duplex switched Ethernet}
\newacronym{CAN}{CAN}{Controller Area Network}

\newacronym{OCI}{OCI}{Open Carrier Interface }

\newacronym{CBR}{CBR}{constant bit rate}

\newacronym{LSTM}{LSTM}{Long Short-Term Memory}

\newacronym{OVS}{OVS}{Open vSwitch}
\newacronym{HTB}{HTB}{Hierarchical Token Bucket}
\newacronym{TRTCM}{TRTCM}{Two-Rate, Three Color Marker}

\usepackage{subcaption}

\usepackage{siunitx} 

\usepackage[noadjust]{cite}

\usepackage[font={footnotesize}]{caption}

\usepackage{textcomp}

\makeatletter
\newcommand*{\inlineequation}[2][]{%
  \begingroup
    \refstepcounter{equation}%
    \ifx\\#1\\%
    \else
      \label{#1}%
    \fi
    \relpenalty=10000 %
    \binoppenalty=10000 %
    \ensuremath{%
      #2%
    }%
    ~\@eqnnum
  \endgroup
}
\makeatother

\ifCLASSINFOpdf
  \usepackage[pdftex]{graphicx}
\else
\fi

\begin{document}
%
% paper title
% Titles are generally capitalized except for words such as a, an, and, as,
% at, but, by, for, in, nor, of, on, or, the, to and up, which are usually
% not capitalized unless they are the first or last word of the title.
% Linebreaks \\ can be used within to get better formatting as desired.
% Do not put math or special symbols in the title.

%\title{Performance Evaluation of Virtual Switching Systems for Real-Time Communication}
\title{Predictable Bandwidth Slicing with Open vSwitch}
%\title{Impact of Hardware Configuration on the Switching Delay of OVS-DPDK}

%\title{Impact of Hardware Configuration on the Switching Delay of OVS-DPDK}

%
%
% author names and IEEE memberships
% note positions of commas and nonbreaking spaces ( ~ ) LaTeX will not break
% a structure at a ~ so this keeps an author's name from being broken across
% two lines.
% use \thanks{} to gain access to the first footnote area
% a separate \thanks must be used for each paragraph as LaTeX2e's \thanks
% was not built to handle multiple paragraphs
%

\author{\IEEEauthorblockN{Jesse Chen and Behnam Dezfouli}
\IEEEauthorblockA{Internet of Things Research Lab, Department of Computer Science and Engineering, Santa Clara University, USA}
\texttt{\small{\{jschen, bdezfouli\}@scu.edu}}
}

\maketitle

% As a general rule, do not put math, special symbols or citations
% in the abstract or keywords.
\begin{abstract}
Software switching, a.k.a virtual switching, plays a vital role in network virtualization and network function virtualization, enhances configurability, and reduces deployment and operational costs.
Software switching also facilitates the development of edge and fog computing networks by allowing the use of commodity hardware for both data processing and packet switching. 
Despite these benefits, characterizing and ensuring deterministic performance with software switches is harder, compared to physical switching appliances.
In particular, achieving deterministic performance is essential to adopt software switching in mission-critical applications, especially those deployed in edge and fog computing architectures.
In this paper, we study the impact of switch configurations on bandwidth slicing and predictable packet latency.
We demonstrate that latency and predictability are dependent on the implementation of the bandwidth slicing mechanism and that the packet schedulers used in OVS Kernel-Path and OVS-DPDK each focus on different aspects of switching performance. 

%Specifically, we show that bandwidth-sliced queues in a software switch can be accurately modeled as an M/M/1 queueing system, and that increasing switch configuration complexity comes with predictable latency increases. 

\end{abstract}


% For peer review papers, you can put extra information on the cover
% page as needed:
% \ifCLASSOPTIONpeerreview
% \begin{center} \bfseries EDICS Category: 3-BBND \end{center}
% \fi
%
% For peerreview papers, this IEEEtran command inserts a page break and
% creates the second title. It will be ignored for other modes.
\IEEEpeerreviewmaketitle

\section{Introduction}
%Cloud computing is the modern structure around which services and applications are built and deployed within limited time and under resource constraints \cite{nist_cloud}. 
%Cloud computing service models \cite{stallings} such as Software as a Service (SaaS), Platform as a Service (PaaS) and Infrastructure as a Service (IaaS), depend on widespread network access availability (either mobile or wired) to shared, scalable and automatically provisioned computing resources. 

With the arrival of new paradigms such as edge and fog computing, the necessity for comprehensive understanding of network behavior becomes increasingly important as tasks often have a multitude of requirements that depend on network performance guarantees, such as minimum flow bandwidth or packet latency constraints. 
%Fulfilling some requirements is easy to accomplish. 
Some requirements are easy to fulfill. 
For example, it is straightforward to track the available processing and memory resources of edge and fog nodes. 
%and allocate these resources to tasks in a deterministic manner.
However, the prediction of network parameters such as end-to-end packet latency is dependent on a variety of factors and requires a comprehensive understanding of network configuration and topology \cite{emmerich2018throughput, javed2017stochastic}.
An essential step towards this understanding is the characterization of packet switching behaviors.

%% why software switches
In this paper, we focus on software switching considering its high applicability in edge and fog computing scenarios \cite{powell2020fog}. 
For example, software switches are utilized to build multi-function nodes in edge and fog systems, where each node performs both networking and computation tasks.
Specifically, with commodity hardware that is capable of computation, software switches are used to add switching capability to a network, resulting in lower costs for deployment, maintenance, and upgrades.
Software switches also offer greater configuration flexibility, making them more suitable for edge and fog networks which need to handle highly-dynamic workloads. 
In contrast, traditional hardware switches, even when enabled with \gls{SDN} protocols such as OpenFlow and NETCONF, are limited in their ability to accomplish effective bandwidth slicing because of queue limitations. 
While hardware switches are usually limited to eight queues per port, software switches do not impose this limitation \cite{heise2015deterministic}. 
Furthermore, the size of software switches' flow tables is flexible and can be extended in ways that hardware switches' can not.
Software switches open up the possibility of efficient bandwidth isolation for each task's data flows and simplify the process of developing and applying new network policies \cite{powell2020fog,sheth2019enhancing}.

Although one of the main benefits of software switches when compared to traditional hardware switches is the high degree of flexibility offered, there remains areas of study that have largely been neglected in existing analyses.  
Specifically, existing studies overwhelmingly focus on the switches' maximum throughput capabilities \cite{fang2018evaluating, mcguinness2018evaluating, meyer2014validated} or latency measurements in best-case, non-realistic scenarios \cite{zhang2019benchmarking, emmerich2018throughput, shanmugalingam2016dpdk, begin2018accurate, sattar2017empirical, javed2017stochastic, manggala2015performance}. 
These studies are important to understand the performance limitations of software switches, but they provide very little to characterize performance in real-world scenarios. 
In particular, these studies fail to provide relevant analysis of packet latency in edge and fog networking scenarios where the bandwidth is sliced to provide queue rate guarantees.

%Our work seeks to identify the configurations and parameters that would be most relevant in an \gls{SDN} enabled fog network and analyze their network performance implications.

In this work, we fill the gap in existing literature by studying how the various aspects of bandwidth slicing such as packet scheduling and queue rate affect latency.
We study and evaluate bandwidth slicing using OVS-Kernel Path (OVS-KP) and OVS-DPDK and identify their strengths and weaknesses in terms of latency and resource efficiency.
In addition to characterizing the latency patterns in bandwidth slicing scenarios, we also identify and analyze the underlying causes of these latency patterns. 
We observe that although the packet latency of OVS-DPDK is considerably lower than that of OVS-KP, the latency of OVS-KP is stable and predictable using M/M/1 queuing.
This is because OVS-KP is able to efficiently utilize the available queue buffers. 
%because of the efficient utilization of available queue buffers by OVS-KP.
OVS-DPDK achieves its lower latency by minimizing the time spent by the packets in the packet scheduler queue; however, this comes at the cost of inefficient resource utilization. 
To keep the queue length short, it drops all packets that are in excess of the allocated queue rate, which results in high TCP retransmsission rates and the need for excess ingress bandwidth in order to maintain the target throughput. 
The observations of this paper can be leveraged to employ software switching in various scenarios, such as for building edge and fog computing systems that need to handle the diverse latency and throughput requirements of IoT applications.

The rest of this paper is organized as follows. 
We present the related work in Section \ref{sec:related_works}.
In Section \ref{sec:methodology}, we overview the two software switches used in this work.
In Section \ref{sec:softswitch_queueing_system}, we discuss the importance and extraction of effective queue rate.
In Section \ref{sec:mm1_latency}, we show that the latency of OVS-KP is predictable using the M/M/1 queueing model.
We discuss the latency of software switching with a user-space data plane in Section \ref{sec:ovs_dpdk_latency}. 
In Section \ref{sec:dpdk_efficiency}, we discuss the resource efficiency tradeoffs between different variants of the Open vSwitch.
In Section \ref{sec:conclusion}, we present discussions on the current and future applicability of this work, highlight its significance, and conclude the paper.

\section{Related Work}
\label{sec:related_works}

Existing works on the performance evaluation of software switches are either limited in scope and ignore latency as a performance parameter, or present an evaluation of oversimplified use-cases that are not representative of real-world network configurations. 

Fang \textit{et al.} \cite{fang2018evaluating} evaluate a broad spectrum of the available software switching solutions and present a direct comparison of the maximum throughput values of each of the evaluated switches. 
Their evaluation of software switches remains surface-level as they focus on the intricacies of inter-switch comparability, leaving much to be desired in terms of performance analysis.  
McGuinness \textit{et al.} \cite{mcguinness2018evaluating} focus on performance evaluations of the BESS software switch in the context of high throughput datacenter use-cases. 
%Their evaluation does include an analysis of rate-limiter performance, but they only evaluate the rate-limiter for accuracy, and neglect to evaluate latency.
Although they evaluate the throughput accuracy of the rate limiter, its effect on latency has been neglected.
Furthermore, datacenters cannot be compared to edge and fog networking scenarios, as the two types of networks have different hardware and applications. 
%Their overall analysis of the BESS software switch once again treats throughput as the focus and does not provide any latency analysis. 
Meyer \textit{et al.} \cite{meyer2014validated} present a model for software switch performance, but limit the scope of their model to only include throughput measurements. 
%
% The studies in \cite{fang2018evaluating, mcguinness2018evaluating, meyer2014validated} 
%In summary, these studies evaluate the performance of software switches, but neglect any measurements of latency and instead focus on throughput characterization.
Overall, these studies evaluate the performance of software switches primarily in terms of throughput and neglect to include any measurements of latency.

In \cite{zhang2019benchmarking}, Zhang \textit{et al.} perform evaluations across various state-of-the-art software switching solutions.
They analyze performance of the OVS-DPDK, snabb, BESS, FastClick, VPP, and netmap VALE software switches and present comparisons of their maximum throughput and packet latency.
Despite the breadth of comparisons, their performance analysis is narrow and only encompasses the most basic of configurations and measurements.  
Emmerich \textit{et al.} \cite{emmerich2018throughput} present an in-depth performance evaluation of \gls{OVS}. 
However, their work primarily focuses on throughput, and the analysis of latency is for very simple scenarios that are insufficient to model edge and fog networking use-cases.  
They evaluate latency only as a function of flow throughput and ignore the performance impacts of bandwidth slicing in multi-queue scenarios.
He \textit{et al.} \cite{he2017low} evaluate a software switch bandwidth slicing mechanism; however, their tests were performed in simple scenarios and their results are presented without a thorough analysis of the latency values.
The same shortcoming is exhibited in \cite{shanmugalingam2016dpdk, begin2018accurate, sattar2017empirical, manggala2015performance} in terms of latency evaluation: their models of packet latency are for simple, synthetic testing scenarios. 
%and do not account for the complexities of real-world software switch deployments.
The latency of a single flow has been evaluated in \cite{shanmugalingam2016dpdk, begin2018accurate, sattar2017empirical}, while \cite{manggala2015performance} only evaluates the latency of a single packet. 
These scenarios are rudimentary and cannot be used to accurately generate models of bandwidth-sliced software switch behaviors. 

\section{Methodology and Background}
\label{sec:methodology}

\begin{figure}[t]
    \centering
    \includegraphics[width=.9\linewidth]{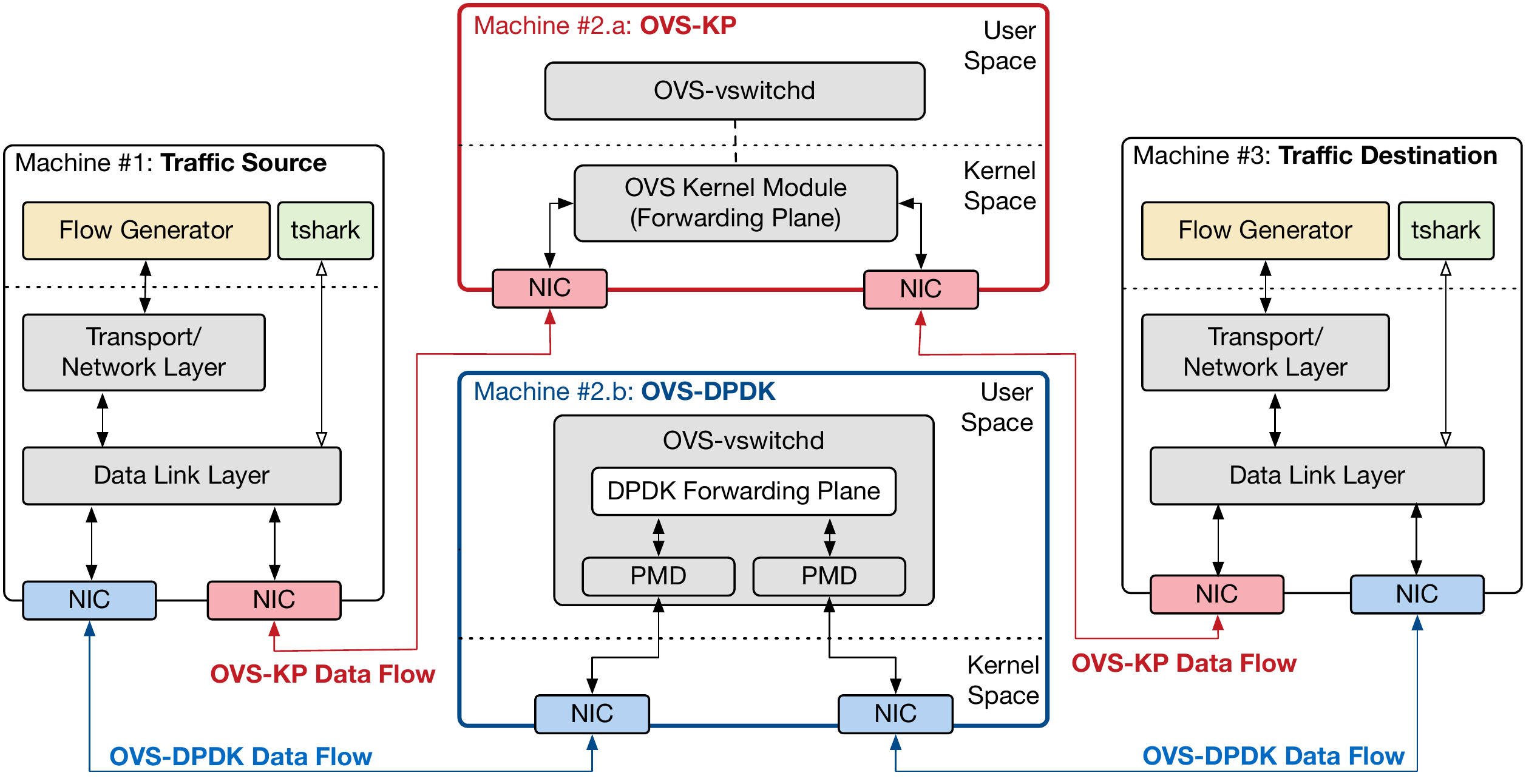}   
    \caption{The testbed used for the studies of this paper. 
    The two software switches used are OVS-KP and OVS-DPDK.
    }
    %In the software switch, we configure our packet scheduler to evaluate various scenarios. }
    \label{fig:test_setup}
\end{figure}

\subsection{Testbed Setup}
%\subsection{Hardware Setup}
Figure \ref{fig:test_setup} presents the testbed architecture. 
To measure the latency of OVS packet switching, we ran experiments on a testbed consisting of three machines: a traffic source, a software switch, and a traffic destination. 
We used Intel 82580 1GbE and Intel X550T 10GbE NICs. 
%The machine on the left is our traffic source VM, and in conjunction with the traffic destination VM on the right, generated all our data flows and flows of interest. 
%In the middle is our \gls{OVS} software switch, which we 
%We then ran experiments with both 1GbE and 10GbE NICs, using Intel 82580 and Intel X550T hardware, respectively. 
% NEED DIAGRAM OF OUR SETUP!!  % WIP 3/26/21
%\subsection{Flow Generation}
%
The traffic source sends UDP and TCP traffic to the traffic destination through the \gls{OVS}. 
%We generated a variety of TCP and UDP flows for our test cases. 
For the UDP flows with fixed bandwidth, we used iPerf, which supports specification of UDP flow bandwidth. 
%For the TCP flows, we use our own flow generator because of iPerf's inability to efficiently generate large numbers of parallel flows. 
%However, for the other tests, where we generate hundreds of concurrent TCP flows, iPerf was not up to the task and we developed our own TCP flow generator, which addresses the shortcomings of iPerf for concurrent flow generation.  
%We modified the switch configurations in different ways for each test.
In the tests, we modified the number of queues, the allocated throughput of each queue, and the type of data flows in each queue. 
We will further discuss the details of each test in their relevant sections.

%\subsection{Packet Capture and Latency Analysis}
Packets are captured using \texttt{tshark} at the egress and ingress ports of the traffic source and traffic destination, respectively.
To ensure the synchronization of timestamp values, the traffic source and traffic destination are two VMs running on a single machine, and the clocks of these two VMs are synchronized with that of the hypervisor.
This configuration allows us to accurately measure and analyze latency values.

\subsection{Open vSwitch}

Open vSwitch (OVS) \cite{pfaff2009extending,pettit2010virtual,pfaff2015design} is an open source, production quality software switch that is compatible with various hypervisors and container systems. 
OVS is highly programmable and is configured using the OpenFlow and OVSDB protocols \cite{chen2021modeling}. 
%such as KVM, Xen, Docker, ESXi, and Hyper-V hypervisors.
%OVS is widely used in cloud computing infrastructures such as OpenStack \cite{sefraoui2012openstack} and Open-Nebula \cite{milojivcic2011opennebula}.
We consider the two main variants of \gls{OVS}:
(i) \gls{OVS} Kernel-Path (\gls{OVS}-KP), which implements its data path via a kernel module, and 
%We refer to this implementation as \textit{OVS-KP}.
(ii) \gls{OVS} with the Data Plane Development Kit (\gls{OVS}-DPDK), which implements its data path through \glspl{PMD} in the user-space.
%Using DPDK, the switch bypasses the kernel and directly accesses the NIC.
%We refer to this implementation as \textit{OVS-DPDK}.
We use \gls{OVS} 2.15.0 and DPDK 20.11.1.

We perform bandwidth slicing on the switches by using  their packet schedulers.
The packet schedulers that we utilize are configured to shape flows via \textit{minimum guaranteed rate} and/or \textit{maximum limited rate} parameters. 
When combined with flow rules that direct packets to the queues, data flows are shaped to specific minimum and maximum rates.

For OVS-KP, we use the \gls{HTB} packet scheduler since it is widely used and available in the Linux traffic control module. 
%HTB is configurable using the OVSDB protocol. 
\gls{HTB} is a classful queuing discipline that supports hierarchical traffic shaping.
Its rate control mechanisms are implemented with the token bucket filter algorithm, and its hierarchical token borrowing system allows parent classes to share tokens with their child classes. 
This token sharing system allows each child class to enforce a guaranteed minimum rate, while also sharing excess available bandwidth with their sibling classes. 
%Although the intricacies and hierarchical token sharing is not the focus of this work, we will use \gls{HTB} to slice flows into separate, rate-limited queues and analyze the resultant latencies. 
%This is accomplished through a token borrowing system that ensures a minimum rate will always be available to a queue while also allowing queue rates to increase towards a maximum rate (burst) if extra bandwidth is available.

% With OVS-DPDK, we use \gls{TRTCM}.
% because OVS-DPDK does not utilize the Linux network stack, and as such does not have access to Linux traffic control's queueing disciplines.  
\gls{OVS}-DPDK uses a different packet scheduler based on the \gls{TRTCM} algorithm. 
Similar to \gls{HTB}, \gls{TRTCM} also uses a token bucket for rate control and provides traffic shaping abilities such as guaranteed minimum and maximum queue rates.

Although HTB and TRTCM are very similar on the surface, their rate control mechanisms are significantly different, which results in different latency and throughput behaviors for queues with the same allocated rate. 
\gls{HTB} is a hierarchical implementation of the token bucket filter algorithm, meaning that when a packet arrives at the head of the queue and there are no tokens available, the packet waits in the queue until tokens become available, at which time the packet is dequeued and sent to the NIC. 
%On the other hand, \gls{TRTCM} serves as a packet marker, and is only responsible for identifying if the packet is within the queue rate limitations. 
On the other hand, for \gls{OVS}-DPDK, when a packet arrives at the head of the queue, the \gls{TRTCM} token buckets are checked for tokens, and if there are not enough tokens for the packet, the packet is dropped. 
Most significantly, this difference in behavior results in different dequeue rates from the queues; as we will discuss in Sections \ref{sec:softswitch_queueing_system} and \ref{sec:ovs_dpdk_latency}, the dequeue rate impacts flow latency and throughput.

% need figure of software switch queueing systems
\section{Extracting Effective Queue Rate}
\label{sec:softswitch_queueing_system}
%To understand which variables can potentially affect the latency of a packet through a software switch, we must first understand the sources of latency. 
In a switching system, a packet experiences three types of latency: transmission latency, processing latency, and queueing latency.  
Transmission latency is directly related to NIC's transmission rate and can be easily calculated. 
%via $L_t = P/R$ where $P$ is the packet size and $R$ is the NIC rate.
%This type of latency is completely dependent on the hardware and will remain constant for all networking hardware that have the same line-rate. 
%Transmission latency is simple to calculate and can be represented as $t_t = L/R$ where $L$ is packet length and $R$ is the hardware line-rate. 
%In this work, we do not focus on transmission latency because it will be the same for both hardware and software switches of the same line-rate. 
%
Processing latency is caused by various factors, including copying a packet to and from different queues, looking up a packet's forwarding decision in the tables, or waiting for hardware I/O operations.  %phrasing is kind of weird here
Since hardware switches are solely dedicated to one function, packet processing delay in a hardware switch is consistently low. 
On the other hand, the processing delay of software switches is usually higher and with greater statistical variation.  
Choice of packet scheduler also affects processing delay. 
Software switches sacrifice processing latency in exchange for greater flexibility. 
Last, and the focus of this section, is \textit{queueing latency} ($L_{q}$), which is defined as the time spent by a packet in the switch's queue, waiting to be transmitted.  
This latency depends on the queue rate, flow rate, and packet scheduler.

\iffalse
\begin{figure}[t]
    \centering
    \includegraphics[width=\linewidth]{IEEE-NETSOFT/figures/softswitch_mechanism/softswitch_queue_org.pdf}   
    \caption{Symbolic representation of a packet's path through a software switch. 
    There are three queues that each packet musht traverse: the ingress queue, the packet scheduler queue, and the egress queue.  }
    \label{fig:softswitch_rep}
\end{figure}

For deeper understanding of our software switching behavior, we first model a software switch as a series of queues that each packet must traverse, then confirm the validity of our representation through empirical observations.
Figure \ref{fig:softswitch_rep} shows the mechanism of queues in a software switch. 
Please note that this is a symbolic representation of a packet's flow within a software switch; ports on a real machine will perform both ingress and egress packet operations. 
As we can see, for each packet that enters the switch, there are three queues that it must traverse: the ingress queue, the egress queue, and the packet scheduler queue.  
\fi

In every software switch, each packet must traverse three queues: the ingress NIC queue, the egress NIC queue, and the packet scheduler's queue. 
%Under normal operating conditions, each NIC will have packets in both its ingress and egress queues. 
%Each of the  queues have different rates, but the limiting factor will be the queue with the lowest throughput. 
%
In this system, throughput is limited by the lowest-rate queue, which is usually the user-allocated packet scheduler queue. 
As a result, a packet spends the most time waiting in the packet scheduler queue because queueing latency experienced by a packet increases exponentially as the queue input rate approaches the queue service rate.
We show that this queueing latency follows an M/M/1 queue trend and that with enough knowledge of the system, the latency can be predicted when using OVS-KP.
%In bandwidth slicing scenarios, the ingress and egress queues are limited by their NICs' line rates, while the queueing discipline queue's rate is limited by the size of the allocated bandwidth slice.
%Each of these queues contribute to the total queueing latency that the packet experiences, but the largest contributor of queueing latency is the bottleneck queue. 

\subsection{Observing Packet Scheduler Behavior}
\label{subsec:qdisc_dequeue}
An important factor in predicting packet latency through any queueing system is the queue's dequeue behavior. 
In our case, we need to understand how the packet schedulers dequeue traffic from their rate-limited queues. 
Although rate-limited queues are allocated with bits/s or bytes/s values, the packet scheduler is not actually dequeueing in such small increments. 
The dequeue behavior varies, depending on how the packet scheduler is implemented, and even different packet schedulers with similar queue parameters behave differently. 
%even different packet schedulers in the same kernel behave differently. 
This often-overlooked variable causes packets to experience different queueing latency values, even if the allocated queue rates are identical. 
% CUT: We performed experiments on \gls{HTB} and \gls{TRTCM} to empirically observe the packet schedulers' dequeue behaviors.  

\begin{figure}
    \begin{subfigure}[t]{0.5\linewidth}
        \centering
        \includegraphics[width=\linewidth]{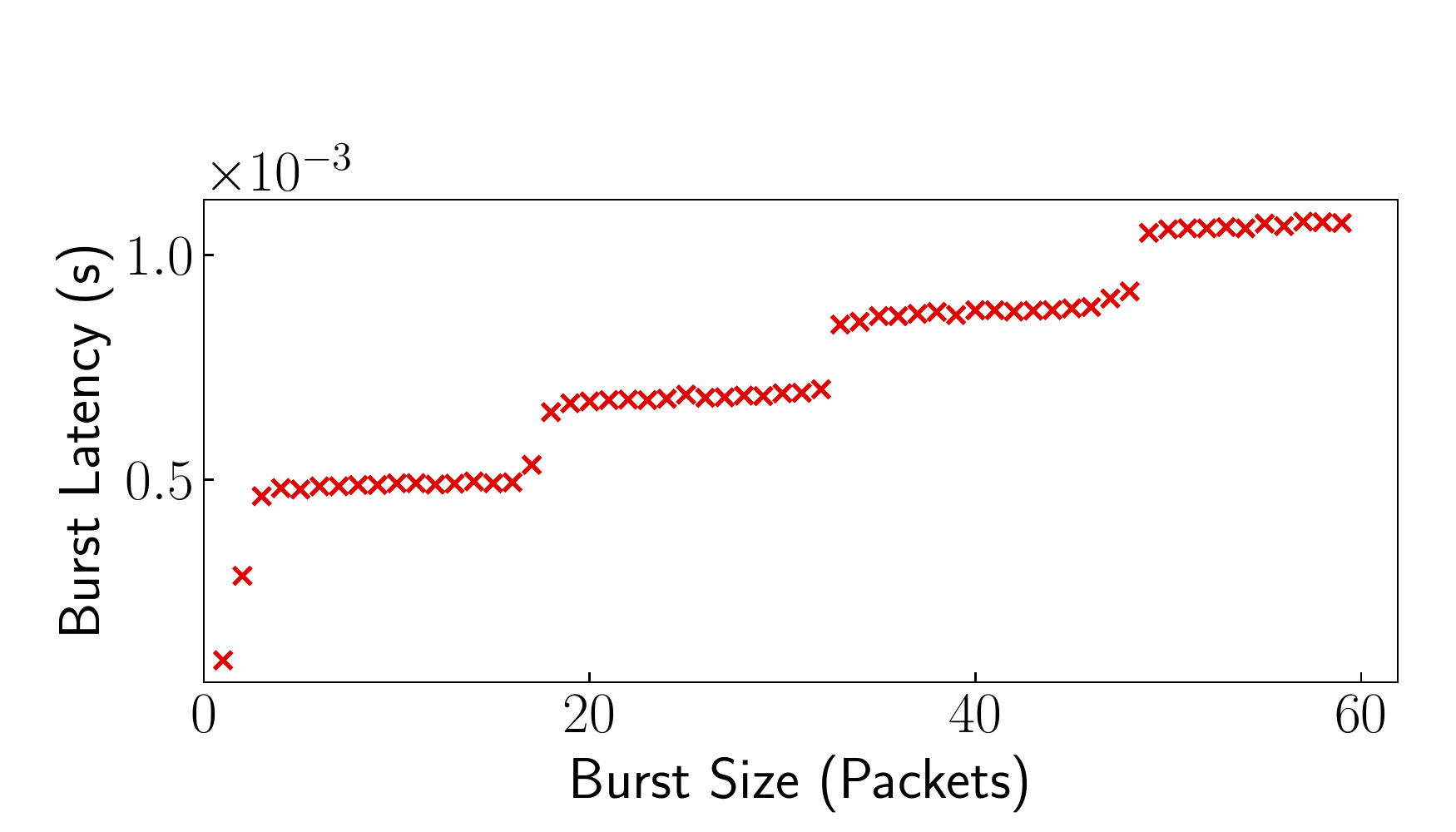}
        \caption{Burst Latency for \gls{OVS}-KP}
        \label{fig:whole_burst_latency-kp}  
    \end{subfigure}%
    \begin{subfigure}[t]{0.5\linewidth}
        \centering
        \includegraphics[width=\linewidth]{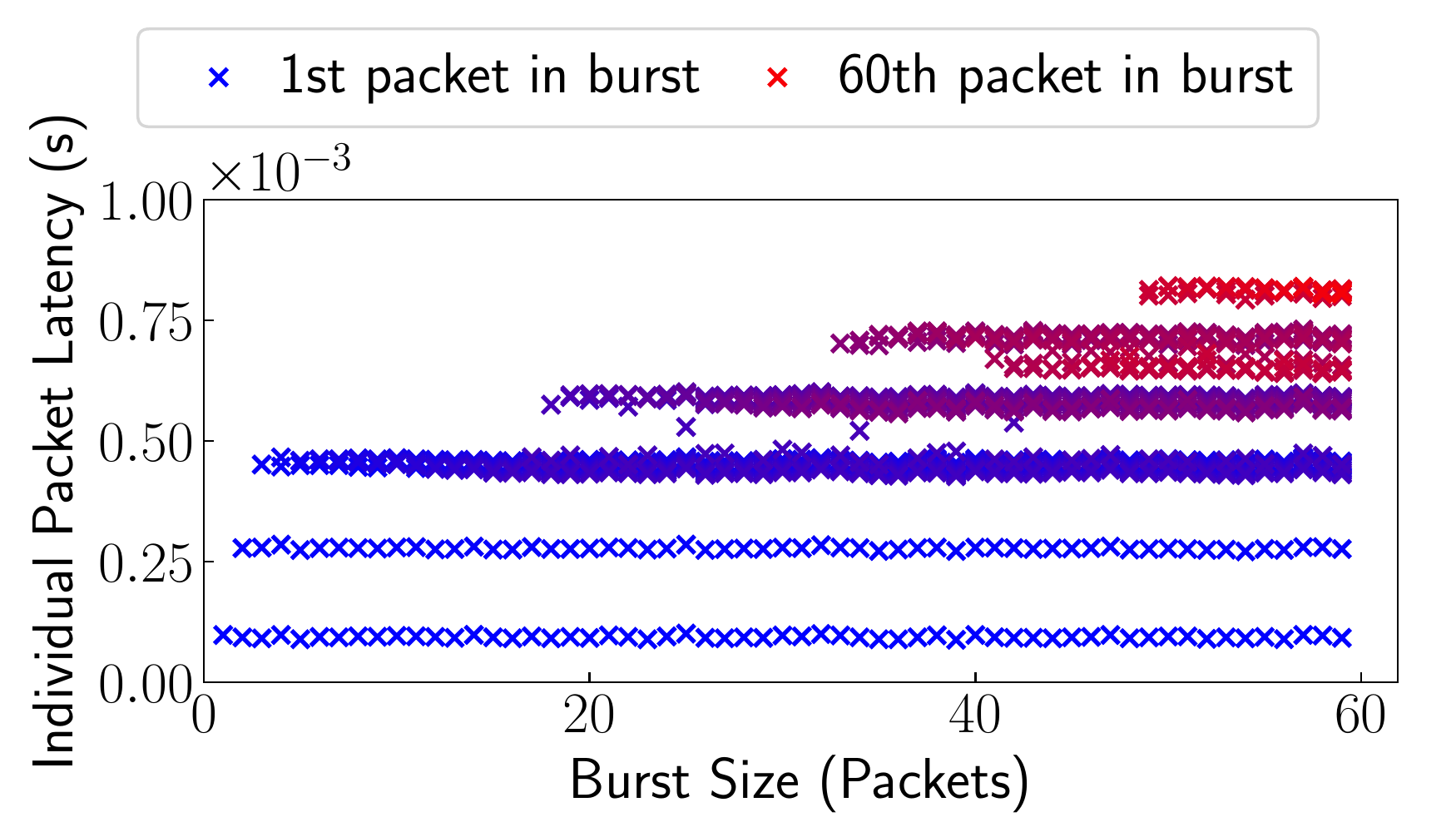}
        \caption{Per Packet Latency for \gls{OVS}-KP}
        \label{fig:per_packet_latency-kp}  
    \end{subfigure}
    \begin{subfigure}[t]{0.5\linewidth}
        \centering
        \includegraphics[width=\linewidth]{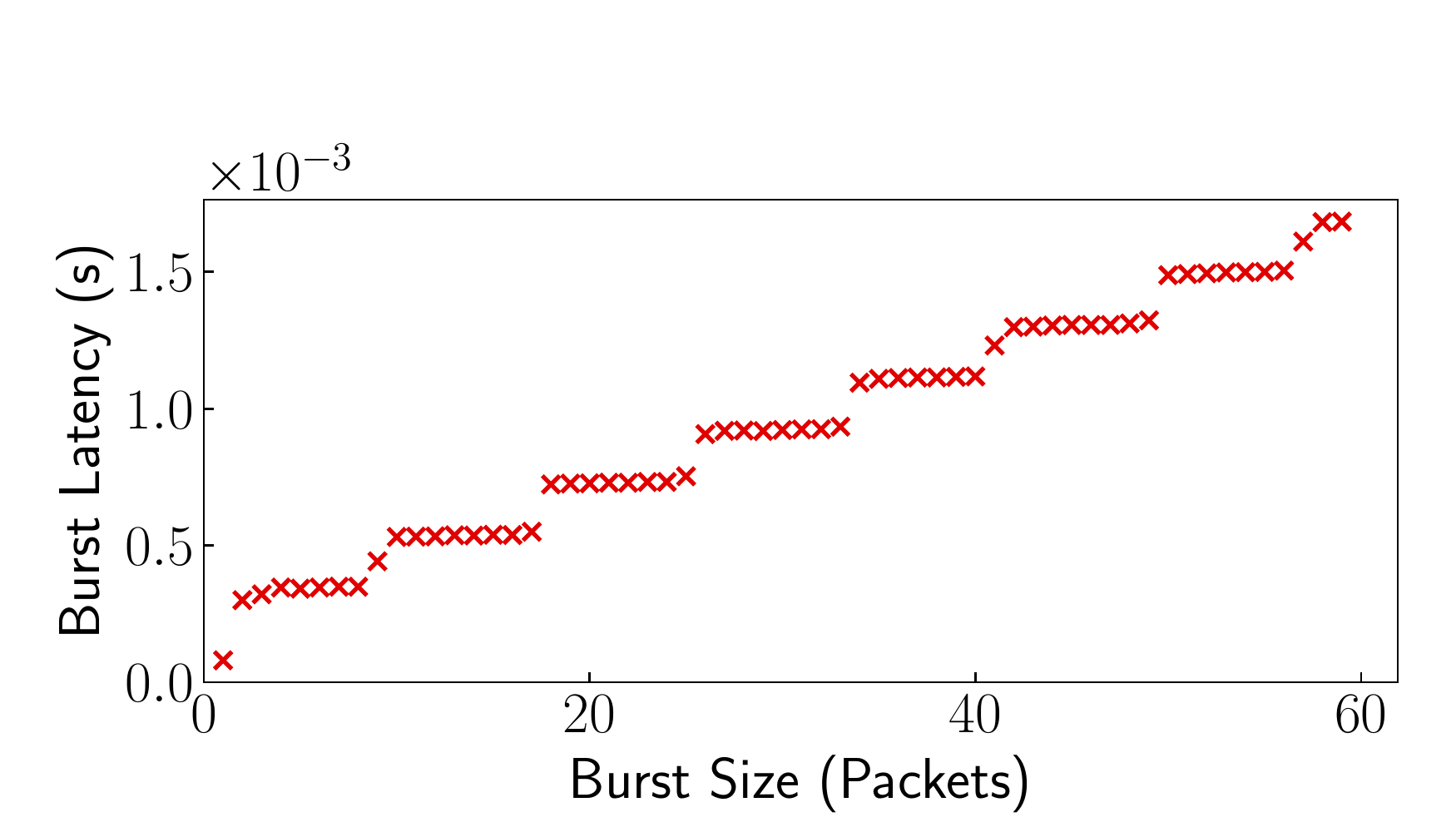}
        \caption{Burst Latency for \gls{OVS}-DPDK}
        \label{fig:whole_burst_latency-dpdk}  
    \end{subfigure}%
    \begin{subfigure}[t]{0.5\linewidth}
        \centering
        \includegraphics[width=\linewidth]{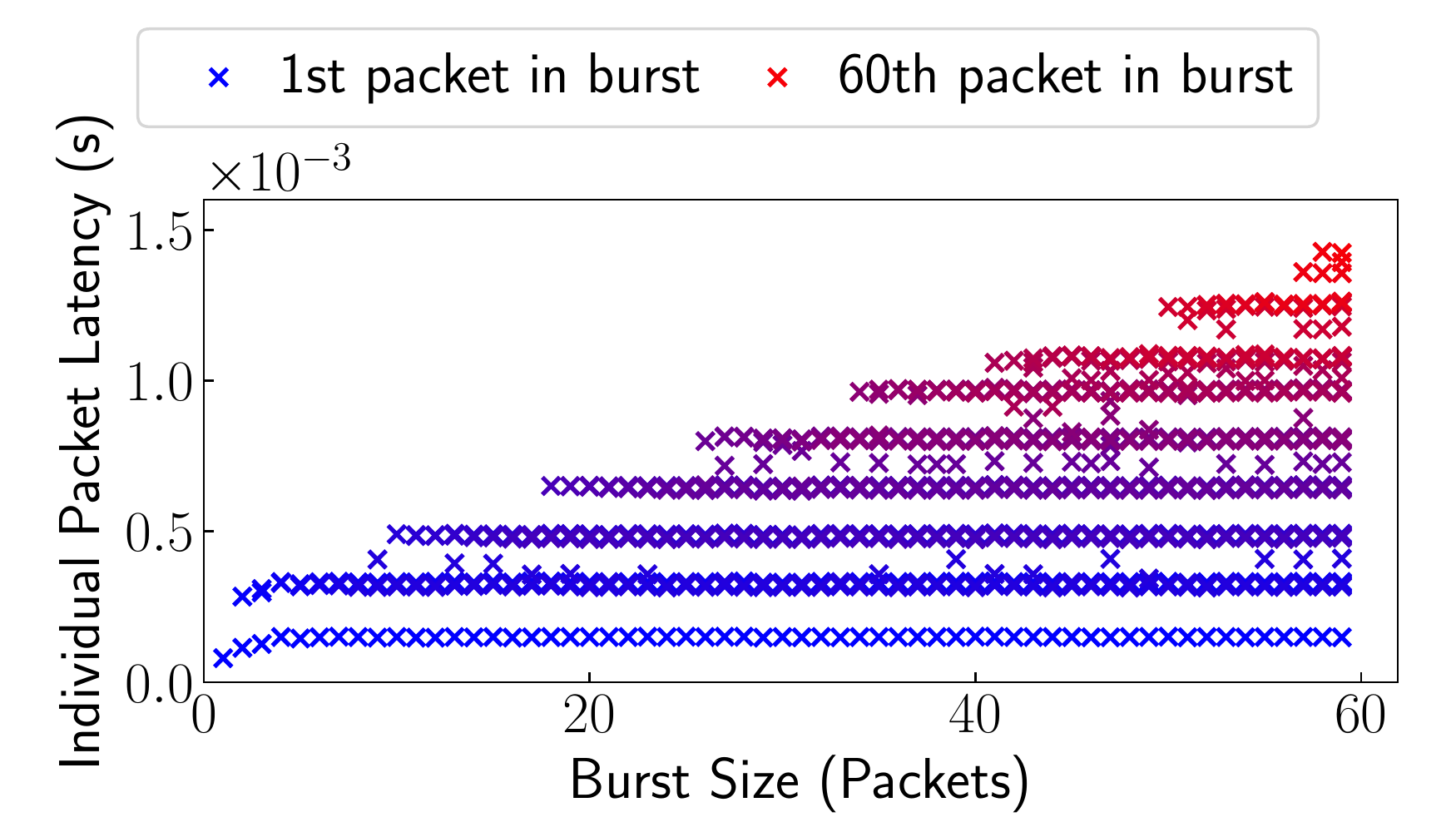}
        \caption{Per Packet Latency for \gls{OVS}-DPDK}
        \label{fig:per_packet_latency-dpdk}  
    \end{subfigure}
    \caption{Packets are dequeued from \gls{HTB} and \gls{TRTCM} queues in different sized groupings. \gls{HTB} dequeues in groupings of 16 packets while \gls{TRTCM} deqeuues in groupings of 8 packets. }
    %This can be observed in both the latency of the entire burst (\ref{fig:whole_burst_latency-kp} and \ref{fig:whole_burst_latency-dpdk}) and the latencies of each individual packet in the burst (\ref{fig:per_packet_latency-kp} and \ref{fig:per_packet_latency-dpdk}). }
    \label{fig:burst_latency}
\end{figure}

In these experiments, we generate UDP flows with various packet burst sizes, then we track the individual packet latencies and the latency of each burst as a whole. 
Figure \ref{fig:burst_latency} presents the results.
As Figures \ref{fig:whole_burst_latency-kp} and \ref{fig:whole_burst_latency-dpdk} demonstrate, when we increase the number of packets in each burst, the latency of the whole burst increases in a stepped pattern. 
This indicates that both \gls{HTB} and \gls{TRTCM} dequeue packets from their queues in bursts of packets, instead of one at a time. 
%The fact that different sized bursts demonstrate identical latencies reveals that multiple packets are dequeued in a single grouping. 
%specifically, groups of about 16 and 8 packets each for \gls{HTB} and \gls{TRTCM}, respectively.
%
To further support this, the analysis of individual packet latencies in Figures \ref{fig:per_packet_latency-kp} and \ref{fig:per_packet_latency-dpdk} show that the latencies of each packet in the bursts are grouped in segments of about 16 and 8 packets each, although there exist small variation in grouping sizes for each dequeue segment. 
%the overall results reveal that packets are not dequeued one at a time, instead, they are dequeued in definitive groupings. 
Figure \ref{fig:per_packet_latency-kp} and \ref{fig:per_packet_latency-dpdk} also highlight that the grouping pattern holds true for bursts of all sizes: no matter how big the burst, packets are always dequeued in fixed size groups. 
%for example, packets 3 through 19 and 20 through 36 are consistently dequeued together. 
Using this information, we extrapolate that the effective rate of \gls{HTB} queues must be calculated in units of 16 packets, and the effective rate of \gls{TRTCM} queues must be calculated in units of 8 packets.

% how this information can be leveraged

\begin{figure}[t]
    \centering
    \includegraphics[width=0.9\linewidth]{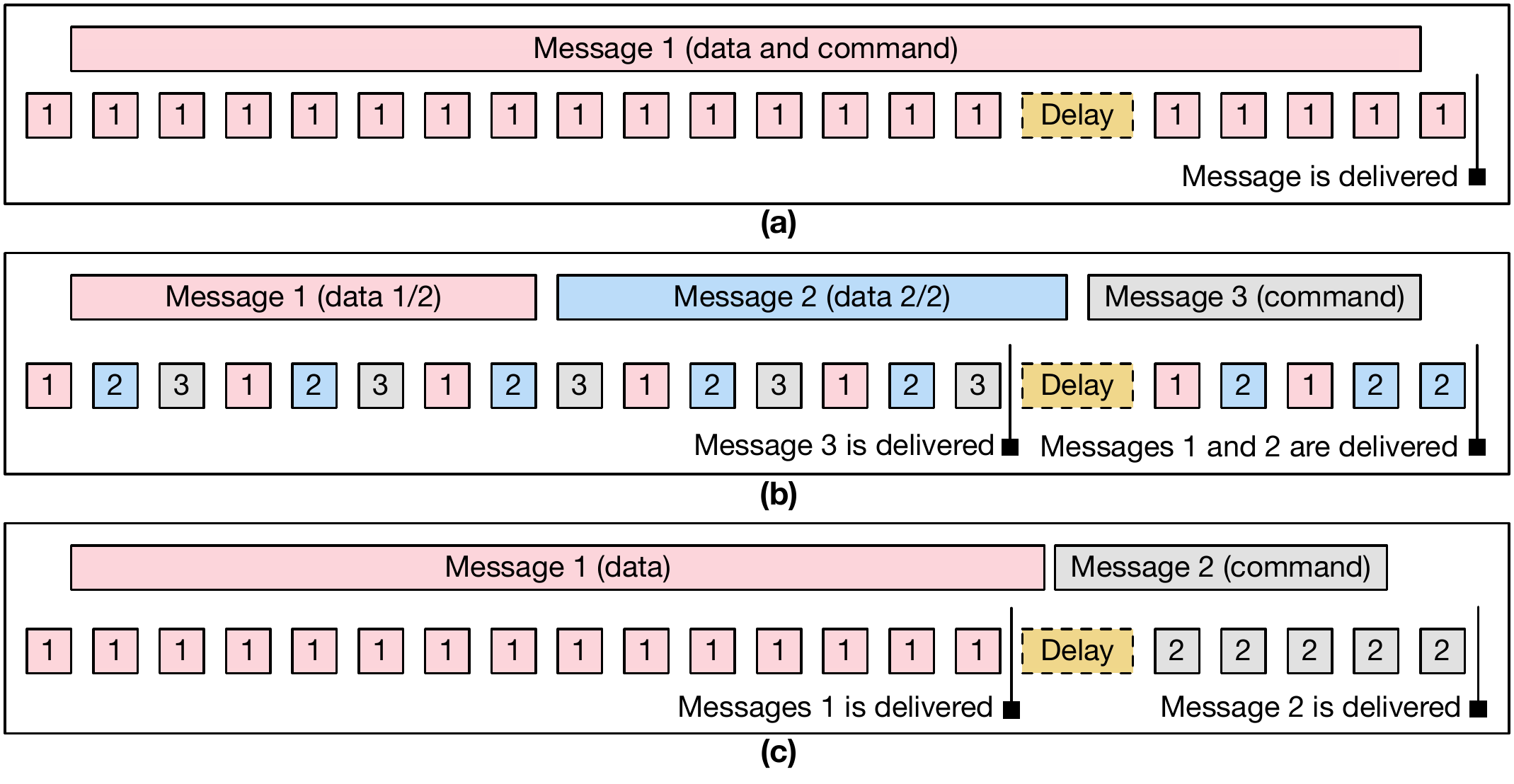}   
    \caption{A scenario where an IoT edge device needs to send data and a command to a server.
    Only case (c) allows faster transmission of data to the server.
    Once the server receives the data, it can start processing, and when the command is received, the action can be performed immediately.
    }
    \label{fig:pkt_size}
\end{figure}

These observations can be leveraged to enhance communication determinism and performance in various contexts.
For example, this knowledge of burst latency behaviors provides devices with the ability to send messages that take advantage of the fact that the latency of a 5-packet message is equal to that of a 15-packet message.
For example, assume an IoT edge device needs to send data and command to a server.
In the first scenario, data and command are sent as a single message, which is segmented into 20 packets, as Figure \ref{fig:pkt_size}a shows.
This results in the delivery of data and command at once.
In the second scenario, two messages are generated for data and one message for command.
Assume the first message is segmented into 7 packets, the second into 8 packets, and the third into 5 packets.
Once these  messages are received by the transport layer of the device, the packets are sent in an interleaved manner.
As it can be observed in Figure \ref{fig:pkt_size}b, the command message is delivered first, which cannot be used because the data messages have not been received yet.
In the third scenario we rely on the behavior of software switches and generate two messages: one for data, which is transmitted first, and one for command, transmitted second.
As Figure \ref{fig:pkt_size}c shows, once the data is received by the server, it can start processing the data, and when the command arrives, the server can perform the action immediately.
Therefore, the third solution provides the minimum latency and better utilization of resources.
A similar method can be used regarding controller-switch communication in \glspl{SDN}.
To ensure timely delivery and execution of commands, the controller can manage the ordering of sent packets based on the command type and size.

\section{Delay Prediction of OVS-KP Bandwidth Slicing}
\label{sec:mm1_latency}
We confirm that the queuing latency of \gls{OVS}-KP switching follows an M/M/1 trend. 
%by showing that NIC ingress and egress queues have a negligible effect on end-to-end packet latency and that the biggest factor in total packet latency is the allocated queue rate.
We set up an experiment to measure the relationship between the latency of packets in a TCP flow and the queue rate. 
% Our setup consisted of three components: a) the traffic source, b) the software switch, and c) the traffic destination and is visualized in Figure \ref{fig:test_setup}. 
We generate and route a TCP flow through a rate-limited queue in the switch.
We do not set any flow rate at the traffic source because the TCP flow rate naturally increases until it detects packet loss caused by the rate-limiter in the software switch.

\begin{figure}
    \begin{subfigure}[t]{0.5\linewidth}
        \centering
        \includegraphics[width=\linewidth]{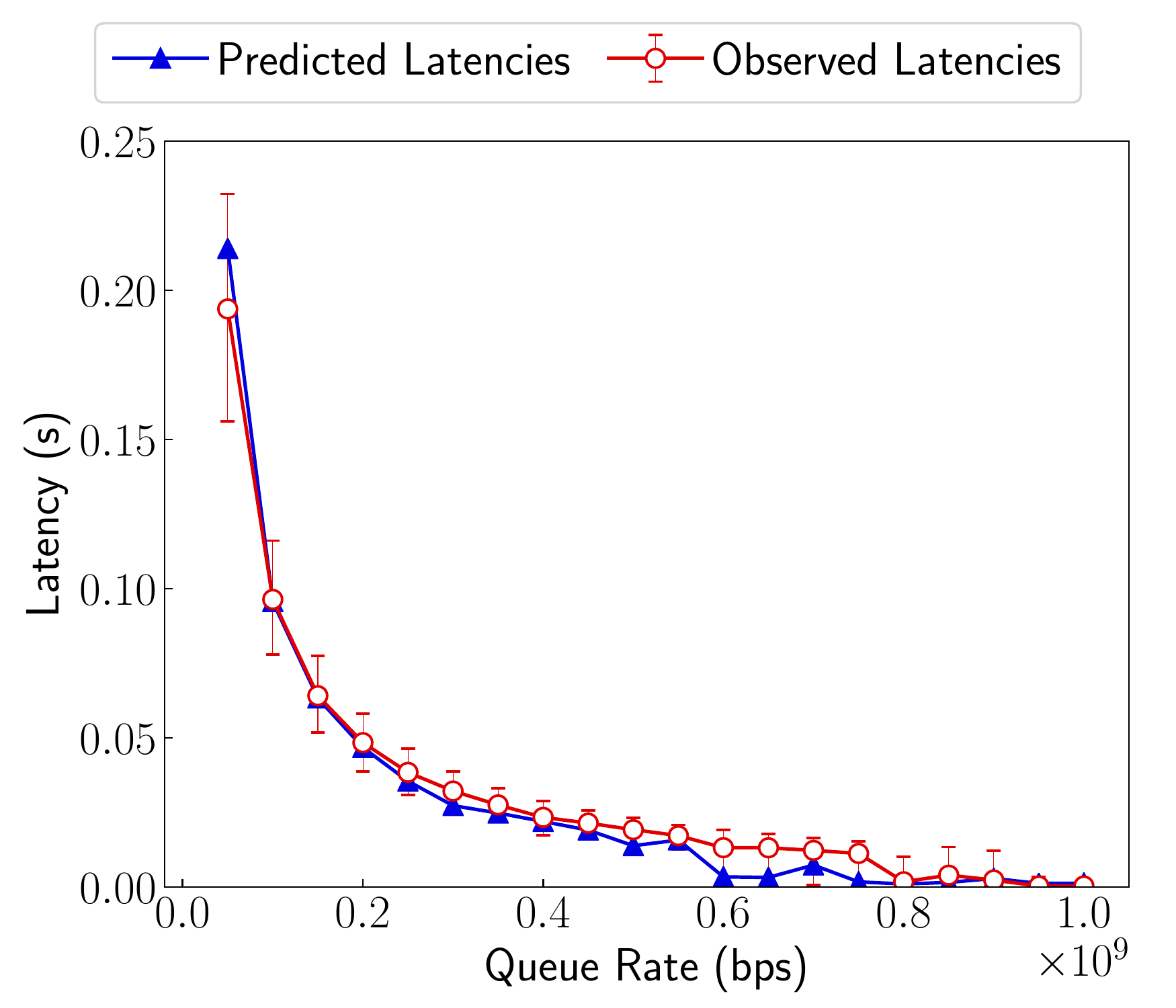}
        \caption{1GbE NIC}
        \label{fig:qrate_latency-1GbE}  
    \end{subfigure}%
    \begin{subfigure}[t]{0.5\linewidth}
        \centering
        \includegraphics[width=\linewidth]{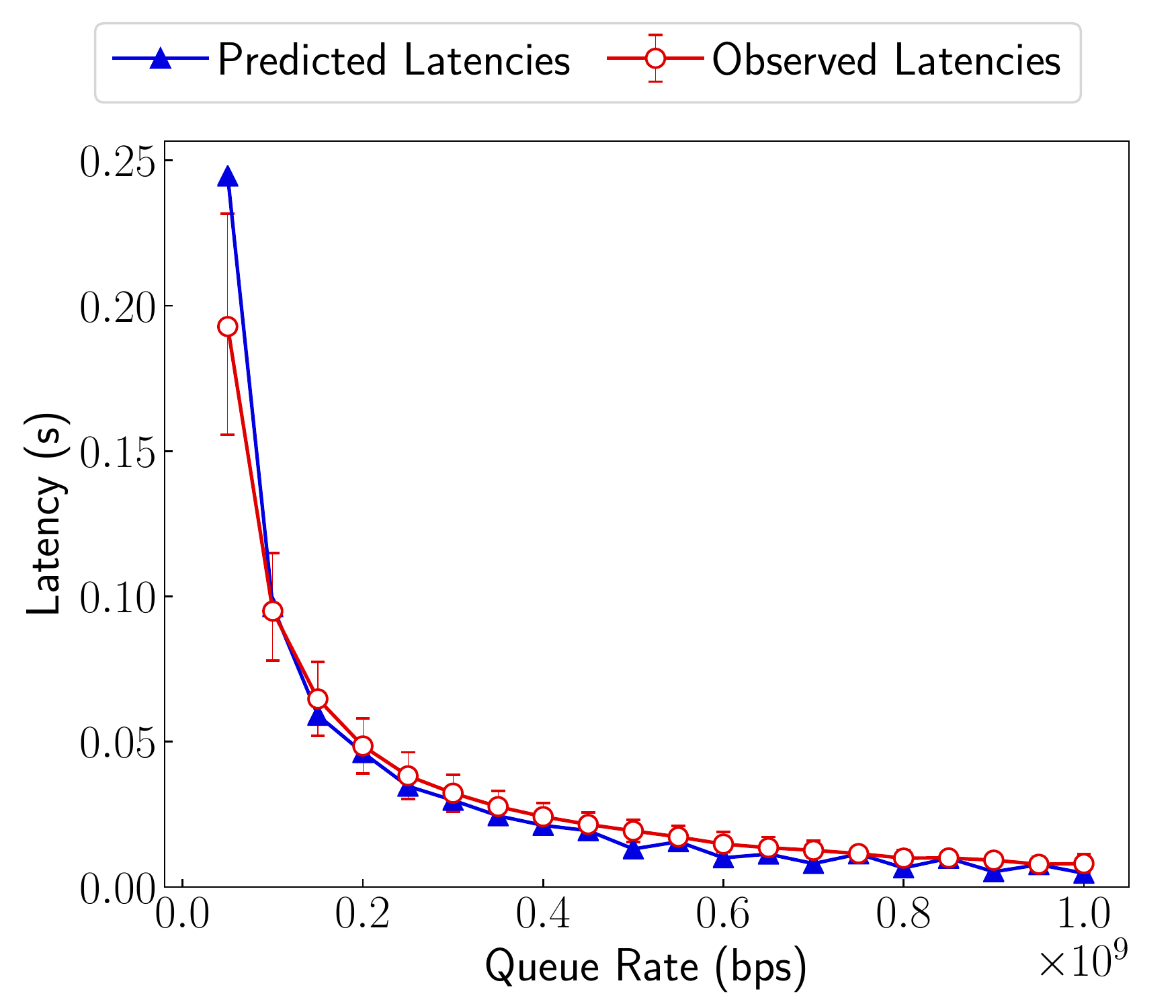}
        \caption{10GbE NIC}
        \label{fig:qrate_latency-10GbE}  
    \end{subfigure}
    \caption{The queueing latency of TCP packets when switched by OVS-KP. 
    The delay is dependent on the allocated queue rate, and with the knowledge of queue ingress rate and packet scheduler implementation, the expected latency can be predicted. }
    \label{fig:qrate_latency}
\end{figure}

We present the results in Figure \ref{fig:qrate_latency}. 
This figure demonstrates that (i) queue rates are the determining factor for rate-limited TCP packet latencies, and (ii) the pattern of observed latencies align with the latency values that one would expect when modeling each queue as an M/M/1 queue. 
The results for 1GbE (Figure \ref{fig:qrate_latency-1GbE}) and 10GbE (Figure \ref{fig:qrate_latency-10GbE}) NICs are presented side-by-side to show that NIC line-rate is not a significant factor in this experiment. 
%and that the value of the queue rates is far more important. 
%Even when queue rates approach the 1GbE line rate and throughput drops a bit, the queue rate is still a far more significant variable than NIC line-rate. 

It is important to note that when calculating the expected latency, the queue rate must be represented via the amount of data that is dequeued in one instance, i.e., the effective queue rate.  
In Section \ref{subsec:qdisc_dequeue}, we showed that \gls{HTB} dequeues 16 packets at a time. 
Thus, instead of calculating expected latency using the queue's bit-rate value, we calculate the expected latency using queue rate in units of 16 packets. 
This is where knowledge of average packet size is important, as we now combine average packet size, queue bit rate, and packet scheduler dequeue behavior to generate effective queue rate values.  
We calculate the expected latency via: \inlineequation[eq:queueing-latency]{L_q = \frac{1}{\mu (1 - \rho)}} where $\mu$ is the effective queue rate, and $\rho$ is the queue utilization ratio \cite{abate1987transient}. 
We include the latency predictions in Figure \ref{fig:qrate_latency} for comparison against the observed values. 
The relationship between queue rate and observed latency is what is expected from an M/M/1 queueing system.
In Figure \ref{fig:qrate_latency}, we validate that predictions based off of observed throughput ($\rho$) and packet scheduler knowledge ($\mu$) are accurate. 
We generate values for $\rho$ by comparing the observed throughput at the traffic source's egress port and the allocated queue rates.
The observed throughput is extracted from the wireshark captures at the traffic source's egress port.
$\mu$ is calculated by converting the units of each queue rate from bits per second to packet groupings per second. 
%The red lines in the figure are the observed latencies, and the blue lines are the latencies that have been predicted from observed traffic. 
%We generate our predicted latencies by calculating each effective queue rate, calculating each queue utilization ratio, and plugging those values into \eqref{eq:queueing_latency}. 
Given that small variations as low as 0.5\% in queue utilization ratio significantly affect latency prediction, the results confirm that this latency prediction methodology is valid and accurate. 

% how this information can be leveraged
%Once again, this information provides additional leverage to \gls{SDN} controllers in edge and fog networks.  
We showed that for bandwidth-sliced flows, queueing latency is the most significant portion of end-to-end latency and that it overshadows transmission latencies; transmission latency on a 1GbE link for a single packet is on the order of microseconds, while the observed switching latencies are up to four orders of magnitude greater. 
Current task allocation schemes either ignore latency as a task request parameter, or assume that network latencies consist only of trivially calculated transmission latencies. 
In contrast, our method allows the prediction of communication latency, which can be leveraged to address the requirements of various tasks in edge and fog computing systems. 
As another example of leveraging this method, a \gls{SDN} controller can accurately enforce bandwidth slicing schemes that satisfy the expected communication latency between edge devices and switches.
Also, to configure switches with latency bounds, the controller can enforce bandwidth slicing methods along all the controller-switch paths.

\section{OVS-DPDK Bandwidth Slicing}
\label{sec:ovs_dpdk_latency}

%%% WRITE DISCUSSION AND ANALYSIS

%In the previous section, we discussed how variations in queueing discipline implementation also significantly affects the queueing latency of a packet through a software switch. 
%In the previous section, we discussed the extent to which effective dequeue unit size impacts a packet's queueing latency. 
In this section, we focus on \gls{OVS}-DPDK and the effect of the \gls{TRTCM} packet scheduler on packet latency, in comparison to \gls{OVS}-KP's \gls{HTB}.
%is on the opposite end of the spectrum when compared to \gls{HTB} with regards to packet latency and resource efficiency. 
%
\begin{figure}  
    \begin{subfigure}[t]{0.5\linewidth}
        \centering
        \includegraphics[width=\linewidth]{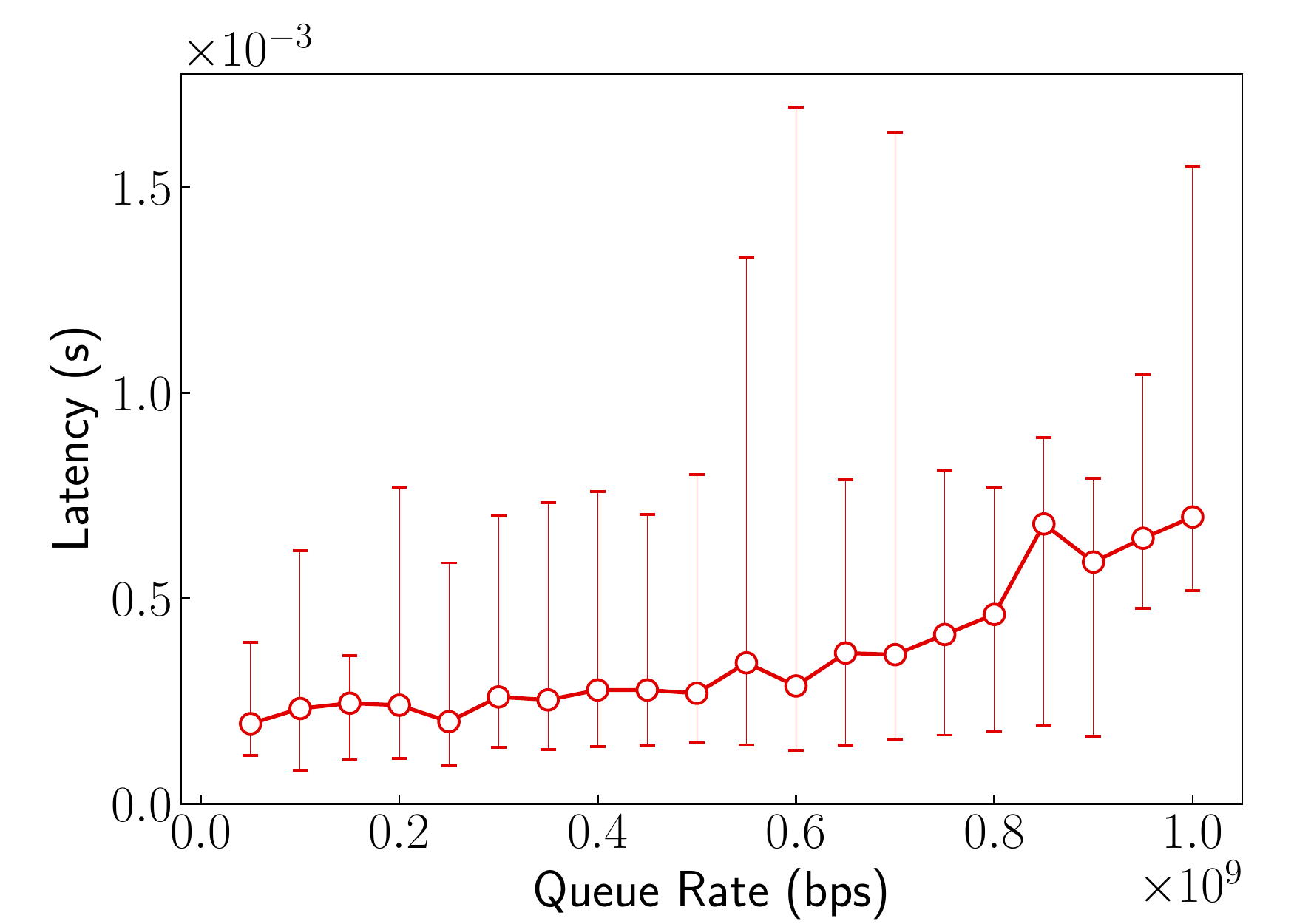}
        \caption{1GbE NIC}
        \label{fig:dpdk_qrate_latency-1GbE}  
    \end{subfigure}%
    \begin{subfigure}[t]{0.5\linewidth}
        \centering
        \includegraphics[width=\linewidth]{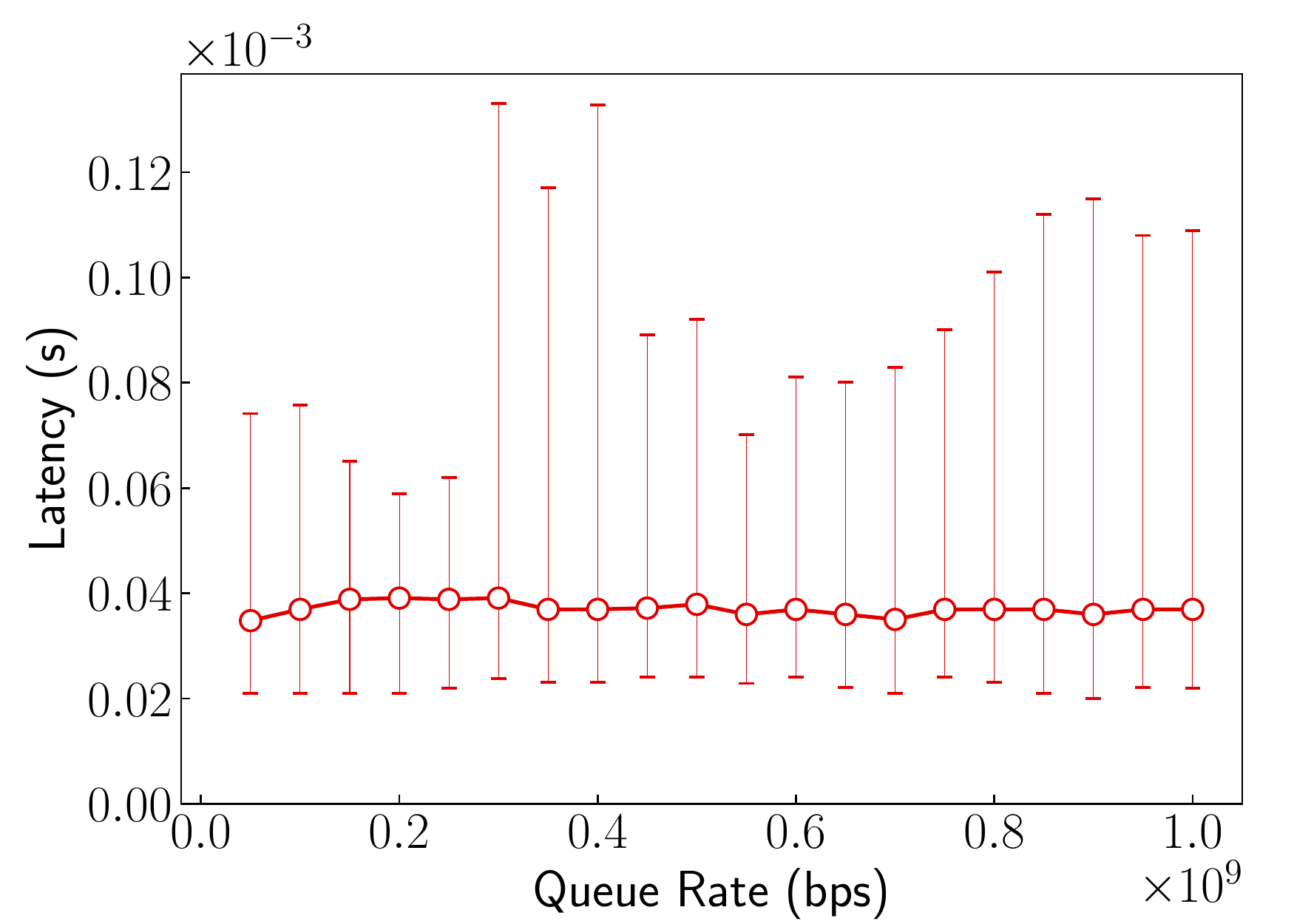}
        \caption{10GbE NIC}
        \label{fig:dpdk_qrate_latency-10GbE}  
    \end{subfigure}
    \caption{For \gls{OVS}-DPDK, the end-to-end packet latency is not as predictable as that of \gls{OVS}-KP due to the differences between \gls{HTB} and \gls{TRTCM}. }
    \label{fig:dpdk_qrate_latency}
\end{figure}
For a direct comparison between the two variants of \gls{OVS}, this time we use \gls{OVS}-DPDK and run an experiment similar to that of Section \ref{sec:mm1_latency}.
We present the results in Figure \ref{fig:dpdk_qrate_latency}. 
The results show that the latency behaviors are not similar at all to that of Figure \ref{fig:qrate_latency}. 
%thereby, \gls{TRTCM} cannot be modeled as an M/M/1 queueing system based on allocated queue rate. 
\gls{OVS}-DPDK queues that are rate-limited with \gls{TRTCM} cannot be modeled as an M/M/1 queueing system because the queues are not being dequeued at the allocated queue rate.
Although the rate of data sent to the egress NIC matches the allocated rate, the rate at which packets are removed from the queue depends on the CPU frequency and \gls{OVS}-DPDK's tick rate.
Unlike \gls{HTB}, which uses the availability of tokens to limit the rate at which packets are removed from the queue, \gls{TRTCM} uses the availability of tokens to decide which actions to take. % with all available packets.
If there are tokens available in the bucket when a packet is dequeued, the packet is passed on to the NIC. 
If there are not enough tokens for the packet, the packet is dropped. 
The token bucket is refilled at the allocated queue rate, hence, the amount of data sent to the NIC is limited by that value. 
This approach results in a very high dequeue rate for all \gls{TRTCM} queues, and the effective dequeue rate is on the order of several Gbps. % (\gls{OVS}-DPDK's maximum throughput). 
For \gls{OVS}-KP, the value of $\rho$ in Equation \eqref{eq:queueing-latency} is close to 1 because the flow rate is approaching the effective queue rate, whereas for \gls{OVS}-DPDK, that value is now much closer to 0 because the effective queue rate is much higher than the flow rate.
This results in packets spending significantly less time waiting in the packet scheduler's queues. 
As we can see from a direct comparison of Figures \ref{fig:qrate_latency-1GbE} and \ref{fig:dpdk_qrate_latency-1GbE}, for TCP flows that are rate limited to 500 Mbps, the latency is reduced from 19.22 ms with \gls{HTB} to 0.27 ms with \gls{TRTCM}, a 70x reduction. 
Although the magnitude of latency reduction varies depending on the allocated queue rate and NIC line rate, this shows that a significant portion of the latency experienced by the packets that traverse \gls{OVS}-KP's rate-limited queues is the time spent waiting in the packet scheduler's queue. 
\gls{OVS}-DPDK's rate control mechanism is able to avoid these long queueing latencies, while still being able to accurately rate-limit the traffic to the egress NIC.

\section{Resource Efficiency Comparisons}
\label{sec:dpdk_efficiency}

On the surface,  \gls{OVS}-KP and \gls{OVS}-DPDK's rate control mechanisms accomplish the same goal: limit the rate of traffic that is sent to the egress NIC.  
Although differences in implementation have significant implications for latency, another implication that is just as important is the effect of the packet scheduler on resource utilization efficiency. 
One of the main goals in edge/fog task allocation is to utilize resources effectively and efficiently, which, for network resources, is usually accomplished through bandwidth slicing and rate control mechanisms.
We have observed that the \gls{HTB} and \gls{TRTCM} packet schedulers are capable of accurately rate-limiting a queue; however, our observations also show that \gls{OVS}-DPDK's choice to drop all packets that are in excess of the allocated queue rate is a tradeoff between latency and effective bandwidth utilization. 

\begin{figure}[t]
    \centering
    \includegraphics[width=\linewidth]{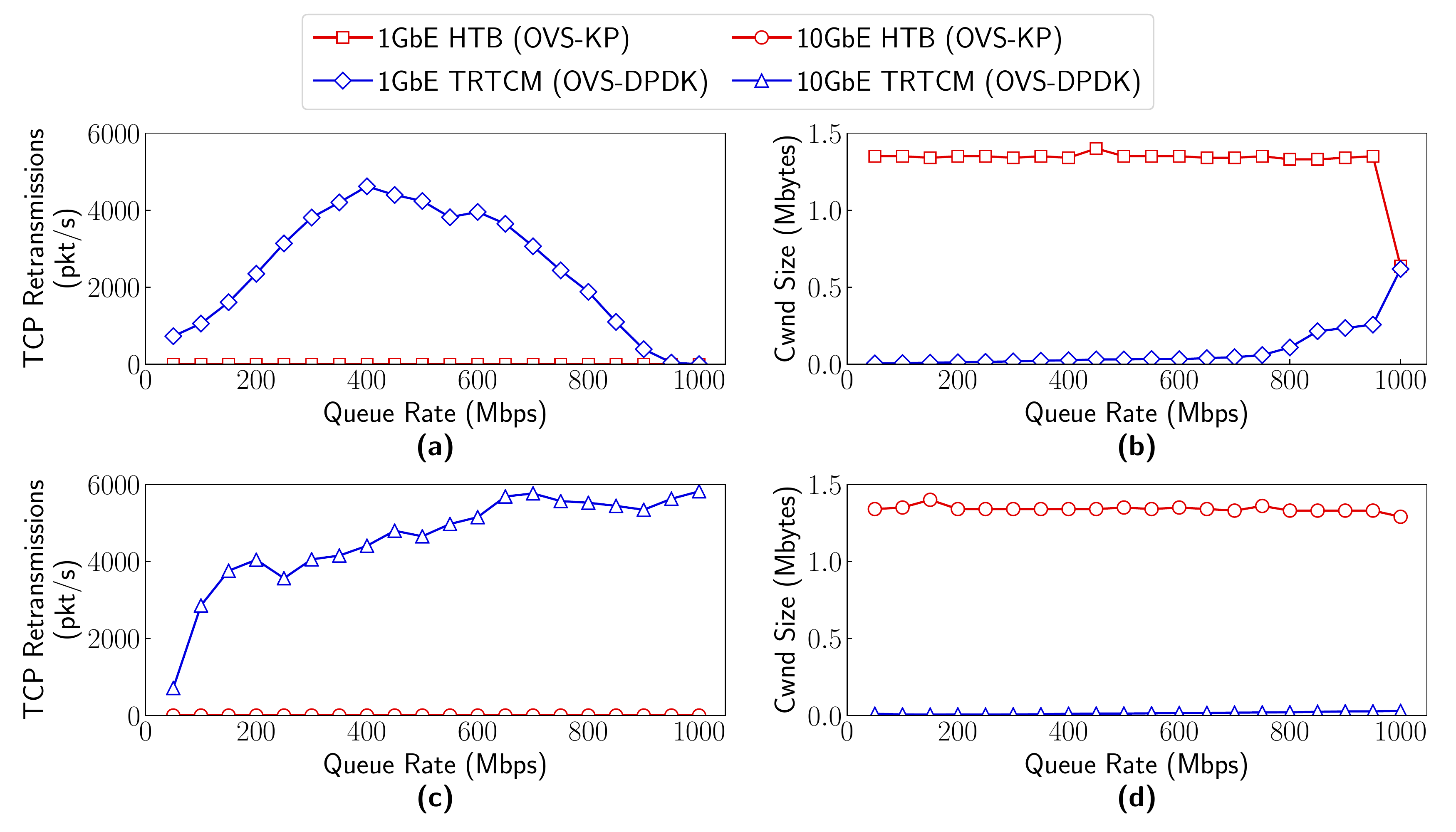}   
    \caption{Comparing the performance of switching a TCP flow through OVS-KP and OVS-DPDK using 1GbE ((a) and (b)) and 10GbE ((c) and (d)) NICs. 
    The queue rate control method of OVS-DPDK is considerably less efficient than that of OVS-KP.}
    \label{fig:iperf}
\end{figure}

We run experiments with the same setup as Figure \ref{fig:qrate_latency} and \ref{fig:dpdk_qrate_latency}, and we use   \texttt{iperf3} to capture data for TCP retransmission rate and TCP congestion window size. 
%We present the results for both 1GbE and 10GbE NICs in Figure \ref{fig:iperf}.
%    
Figures \ref{fig:iperf}a and \ref{fig:iperf}c present the results for TCP retransmission rate. 
We observe that rate-limiting with \gls{TRTCM} causes significantly more TCP retransmissions compared to \gls{HTB}.
%\gls{HTB} rate-limiting results in zero retransmissions across all the tests, while \gls{TRTCM} rate-limiting results in thousands of retransmissions per second.
Rate-limiting to 500 Mbps with \gls{TRTCM} results in 4233 and 4652 TCP retransmissions per second for 1GbE and 10GbE NICs, respectively.
This indicates that for TCP traffic, maintaining an egress throughput of 500 Mbps out of the switch requires an additional 50.8 Mbps and 55.8 Mbps of retransmission traffic, due to the large number of packets that \gls{TRTCM} drops. 
%In the case of 500 Mbps rate-limited queues, \gls{TRTCM} drops 10\% of its packets and requires excess ingress traffic to maintain the allocated egress rate. 
%\gls{OVS}-DPDK does not utilize its queue buffers effectively and as a result, is inefficient in its usage of network resources. 
%
More importantly, although \gls{OVS}-DPDK is able to switch individual packets with lower latency than \gls{OVS}-KP, the high rate of packet drops/retransmissions has an adverse effect on application message latency.
The application layer is not dependent on individual packet latency, rather, is dependent on the latency of messages which can be composed of multiple packets. 
In a situation with a 10\% retransmission rate, large application layer messages are very likely to experience retransmissions and slowdowns due to the inefficiencies of \gls{TRTCM}. 

Figures \ref{fig:iperf}b and \ref{fig:iperf}d show that rate-limited flows with \gls{TRTCM} have much smaller congestion window sizes than flows that are rate-limited with \gls{HTB}. 
Once again, this is related to \gls{TRTCM}---dropping all packets that are over the allocated queue rate is extremely limiting for TCP congestion window size. 
Each time a packet is dropped and retransmitted, the congestion window of that TCP connection is halved. 
For TCP flows with high retransmission rates like those we observed with \gls{TRTCM}, the congestion windows are severely limited and are unable to grow due to the constant packet drops and subsequent window size adjustments. 
The frequent congestion window size adjustments also results in spikes and dips in flow throughput, which have a detrimental effect on latency predictability.  
As such, application layer messages that are sent through \gls{OVS}-DPDK always have an element of unpredictability due to high retransmission rates while messages sent through \gls{OVS}-KP do not.

Since \gls{OVS}-DPDK operates completely in user-space, it achieves its high performance by constantly consuming 100\% of at least one processor core.
For high performance use-cases, a separate core is used for each port, resulting in several processor cores being dedicated entirely to running the DPDK user-space data path. 
In low-cost and low-energy edge and fog computing scenarios, this is not desirable, especially when compared to \gls{OVS}-KP, which consumes less than 5\% of a single processor core with \gls{HTB} while switching 10Gbps traffic with hundreds of flow rules and queues. 

%This is another tradeoff that \gls{OVS}-DPDK makes in its implementation of \gls{TRTCM}: it trades low latency for inefficient resource usage and performance metric instabilities. 

% retransmissions: 
    %- for trtcm 1GbE this results in up to 10% of all traffic is retranmissions, leading to 90% effective bandwidth utilization
    %- for trtcm 10GbE, 

%%% our observations of resource efficiency tell a different 

\section{Conclusion and Future Work}
\label{sec:conclusion}

%The first step towards utilizing software switches in an effective manner is identifying and modeling the variables that affect their performance. 
In this paper, we studied how packet schedulers affect switching latency and resource efficiency. %
We first developed models to predict the latency of the M/M/1 queueing system that can be found in the \gls{HTB} packet scheduler. Specifically, we analyzed the behavior of the packet scheduler, then used these observations to predict packet latency of TCP flows. 
%Through empirical observations of packet switching latency, we demonstrated that queueing latency in a software switch can be predicted by modeling rate-limited queues as an M/M/1 queueing system. 
%We also highlighted the importance of queue rate representation accuracy for packet latency predictions; the effective queue rate used to predict packet latency is dependent on packet scheduler implementation and can be variable between different systems. 
We then discussed the design differences between \gls{OVS}-KP and \gls{OVS}-DPDK packet schedulers and showed how each achieves either low latency or resource utilization efficiency at the cost of the other. 
%Finally, we presented ideas regarding the applicability of each packet scheduler, and highlighted directions for the development of future works. 
The results presented in this work provide a foundation from which we can begin to build deterministic software switching systems that can be specifically used to build low-cost processing and packet switching systems using commodity hardware.

The design decisions that allow \gls{OVS}-DPDK's \gls{TRTCM} to achieve low latency in comparison to \gls{OVS}-KP's \gls{HTB} come at the cost of inefficient bandwidth usage, throughput instability, and reduced latency predictability. %which accomplishes bandwidth slicing via the \gls{HTB} packet scheduler.
This information is especially important to design networks with specific performance metrics in mind. 
Besides, this information can be leveraged to design packet schedulers that combine the desired properties of \gls{HTB} and \gls{TRTCM}. %can be designed and implemented. 
For example, a new packet scheduler that seeks to enforce latency bounds while also achieving flow reliability could dequeue packets according to the queue rate similar to \gls{HTB}, and also dynamically adjust queue length so that packets that do not meet the packet latency requirements are dropped, similar to \gls{TRTCM}. 
This way, the benefits of using the queue buffers can be realized, while also keeping queueing latency within established bounds.

\ifCLASSOPTIONcaptionsoff
  \newpage
\fi

\bibliographystyle{IEEEtran}
\bibliography{refs}

% that's all folks
\end{document}